\documentstyle[sprocl]{article}
\input psfig
\def\beq{\begin{eqnarray}}
\def\eeq{\end{eqnarray}}
\def\bea{\begin{eqnarray*}}
\def\eea{\end{eqnarray*}}

\def\centeron#1#2{{\setbox0=\hbox{#1}\setbox1=\hbox{#2}\ifdim
\wd1>\wd0\kern.5\wd1\kern-.5\wd0\fi
\copy0\kern-.5\wd0\kern-.5\wd1\copy1\ifdim\wd0>\wd1
\kern.5\wd0\kern-.5\wd1\fi}}
\def\ltap{\;\centeron{\raise.35ex\hbox{$<$}}{\lower.65ex\hbox{$\sim$}}\;}
\def\gtap{\;\centeron{\raise.35ex\hbox{$>$}}{\lower.65ex\hbox{$\sim$}}\;}

\def\singleandthirdspaced{\baselineskip=\normalbaselineskip\multiply
    \baselineskip by 130\divide\baselineskip by 100}
\def\singlespaced{\baselineskip=\normalbaselineskip}

\def\dslash{\not{\hbox{\kern-2pt $\partial$}}}
\def\Dslash{\not{\hbox{\kern-4pt $D$}}}
\def\Oslash{\not{\hbox{\kern-4pt $O$}}}
\def\Qslash{\not{\hbox{\kern-4pt $Q$}}}
\def\pslash{\not{\hbox{\kern-2.3pt $p$}}}
\def\kslash{\not{\hbox{\kern-2.3pt $k$}}}
\def\qslash{\not{\hbox{\kern-2.3pt $q$}}}
\def\epsilonslash{\not{\hbox{\kern-2.3pt $\epsilon$}}}

\newcommand{\newc}{\newcommand}
\newc{\qbar}{{\overline q}}
\newc{\Kahler}{K\"ahler }
\newc{\deltaGS}{\delta_{\rm GS}}
\begin{document}
\begin{titlepage}
\begin{flushright}
{\large hep-th/0011376 \\ SCIPP-00/30\\}

\end{flushright}

\vskip 1.2cm

\begin{center}

{\LARGE\bf
TASI Lectures on The Strong CP Problem}

\vskip 1.4cm

{\large Michael Dine}
\\
\vskip 0.4cm
{\it Santa Cruz Institute for Particle Physics,
     Santa Cruz CA 95064  } \\

\vskip 4pt

\vskip 1.5cm

\begin{abstract}
These lectures discuss the $\theta$ parameter of QCD.
After an introduction to anomalies in four and two dimensions,
the parameter is introduced.  That such topological
parameters can have physical effects is illustrated with
two dimensional models, and then explained in QCD using
instantons and current algebra.  Possible solutions
including axions, a massless up quark, and spontaneous
CP violation are discussed.

\end{abstract}

\end{center}

\vskip 1.0 cm

\end{titlepage}
\setcounter{footnote}{0} \setcounter{page}{2}
\setcounter{section}{0} \setcounter{subsection}{0}
\setcounter{subsubsection}{0}

\singleandthirdspaced


\section{Introduction}

Originally, one thought of QCD as being described a gauge coupling
at a particular scale and the quark masses.  But it soon came to
be recognized that the theory has another parameter, the $\theta$
parameter, associated with an additional term in the lagrangian:
\beq
{\cal L} = \theta {1 \over 16 \pi^2} F_{\mu \nu}^a \tilde F^{\mu
\nu a}
\label{thetaterm}
\eeq
where
\beq
\tilde F_{\mu \nu \rho \sigma}^a = {1 \over 2} \epsilon_{\mu \nu
\rho \sigma} F^{\rho \sigma a}.
\eeq
This term, as we will discuss, is a total divergence, and one
might imagine that it is irrelevant to physics, but this is not
the case.  Because the operator violates CP, it can contribute to
the neutron electric dipole moment, $d_n$.  The current
experimental limit sets a strong limit on $\theta$, $\theta \ll
10^{-9}$.  The problem of why $\theta$ is so small is known as the
strong CP problem.  Understanding the problem and its possible solutions
is the subject of this lectures.

In thinking about CP violation in the Standard Model, one usually
starts by counting the parameters of the unitary matrices which
diagonalize the quark and lepton masses, and then counting the
number of possible redefinitions of the quark and lepton fields.
In doing this counting, however, there is a subtlety.  One of
the field redefinitions induces a new term in the lagrangian,
precisely the $\theta$ term above.  The obstruction to making
this transformation is known as an anomaly, and it is not
difficult to understand.

Before considering real QCD, consider a simpler theory, with only
a single flavor of quark.  Before making
any field redefinitions, the lagrangian takes the form: \beq {\cal
L}= -{1 \over 4 g^2} F_{\mu \nu}^2 + \bar q \qslash^* + q \qslash^*
 m \bar q q + m^* \bar q^* q^*.
\eeq
Here, I have written the lagrangian in terms of two-component
fermions, and noted that a priori, the mass need not be real,
\beq
m = \vert m \vert e^{i \theta}.
\eeq
In
terms of four-component fermions, \beq {\cal L} = {\rm Re ~m}~
\bar q q + {\rm Im ~m}~ q \bar q \gamma_5 q. \eeq
In order to bring the mass term to the conventional form, with no
$\gamma_5$'s, one would naively let
\beq
q \rightarrow e^{-i \theta/2} q ~~~~~\bar a \rightarrow e^{-i
\theta/2} \bar q.
\eeq
However, a simple calculation shows that there is a difficulty
associated with the anomaly. Suppose,
first, that $M$ is very large.  In that case we want to integrate
out the quarks and obtain a low energy effective theory.
To do this, we study the path integral:
\beq Z = \int
[dA_{\mu}] \int [dq] [d\bar q] e^{i S} \eeq
Again suppose $m=e^{i \theta} M$, where $\theta$ is small
and $M$ is real.   In order to make $m$ real, we can again make the
transformations:
$q \rightarrow q e^{-i \theta/2};
\bar q \rightarrow \bar q e^{-i \theta/2}$ (in four component
language, this is $q \rightarrow^{-i \theta/2 \gamma_5} q$).)
The result of integrating out the quark, i.e. of performing the
path integral over $q$ and $\bar q$ can be written in the form:
 \beq Z = \int [dA_{\mu}] \int
 e^{iS_{eff}} \eeq
Here $S_{eff}$ is the effective action which describes the
interactions of gluons at scales well below $M$.

\begin{figure}[htbp]
\centering
\centerline{\psfig{file=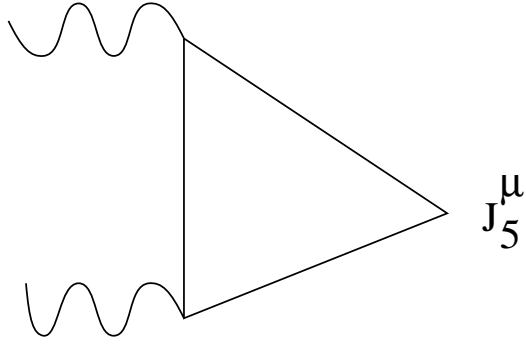,angle=-90,width=7cm}}
\caption{The triangle diagram associated with the four dimensional
anomaly.}
\label{triangle}
\end{figure}

Because the field redefinition which eliminates
$\theta$ is just a change of variables in the path integral,
one might expect that there can be no $\theta$-dependence in the
effective action.  But this is not the case.  To see this, suppose
that $\theta$ is small, and instead of making
the transformation, treat the $\theta$ term as a small
perturbation, and expand the exponential.  Now consider a term in the
effective action with two external gauge bosons.  This is given by
the Feynman diagram in fig. \ref{triangle}.  The corresponding term in the
action is given by
\beq
{\delta}{\cal L}_{eff} = -i {\theta \over 2} g^2  M Tr(T^a T^b)\int {d^4 k \over (2 \pi)^4}
{\rm Tr} \gamma_5 {1 \over \kslash +\qslash_1 - M} \epsilonslash_1
{1 \over \kslash  - M}  \epsilonslash_2
{1 \over \kslash - \qslash_2 - M}
\eeq
Here, as in the figure, the $q_i$'s are the momenta of the two
photons, while the $\epsilon$'s are their polarizations and
$a$ and $b$ are the color indices of the gluons.  To perform the
integral, it is convenient to introduce Feynman parameters and
shift the $k$ integral, giving:
\beq
{\delta}{\cal L}_{eff} = -i \theta g^2  M {\rm Tr}(T^a T^b)
\int d\alpha_1 d \alpha_2 \int {d^4 k \over (2 \pi)^4}
{\rm Tr} \gamma_5 (\kslash - \alpha_1 \qslash_1 + \alpha_2 \qslash_2+\qslash_1 +M)
\epsilonslash_1 \\
{(\kslash - \alpha_1 \qslash_1 + \alpha_2 \qslash_2 +M
)\epsilonslash_2
(\kslash - \alpha_1 \qslash_1 + \alpha_2 \qslash_2 - \qslash_2 + M)
\over (k^2 -M^2 +  O(q_i^2))^3}
\eeq
For small $q$, we can neglect the $q$-dependence of the
denominator.  The trace in the numerator is easy to evaluate,
since we can drop terms linear in $k$.  This gives, after
performing the integrals over the $\alpha$'s,
\beq
{\delta}{\cal L}_{eff}=  g^2 M^2 \theta{\rm Tr}(T^a T^b)  \epsilon_{\mu \nu \rho \sigma}
q_1^{\mu} q_2^{\nu} \epsilon_1^{\rho} \epsilon_2^{\sigma}
\int {d^4 k \over (2 \pi)^4} {1 \over (k^2 -M^2)^3}.
\eeq
This corresponds to a term in the effective action, after doing
the integral over $k$ and including a combinatoric factor of two
from the different ways to contract the gauge bosons:
\beq
{\delta}{\cal L}_{eff}= {1 \over 32 \pi^2} \theta {\rm Tr} (F \tilde
F).
\eeq


Now why does this happen?  At the level of the path
integral, the transformation would seem to be a simple
change of variables, and it is hard to see why this
should have any effect.  On the other hand, if one examines the diagram of fig.
\ref{triangle},
one sees that it contains terms which are linearly divergent, and
thus it should be regulated.  A simple way to regulate the diagram
is to introduce a Pauli-Villars regulator, which means that one
subtracts off a corresponding amplitude with some very large mass
$\Lambda$.  However, we have just seen that the result is
independent of $\Lambda$!  This sort of behavior is characteristic
of an anomaly.

Consider now the case that $m \ll \Lambda_{QCD}$.  In this case, we
shouldn't
integrate out the quarks, but we still need to take into account
the regulator diagrams.  For small $m$, the classical theory has an
approximate symmetry under which
\beq
q \rightarrow e^{i \alpha}q~~~~~~\bar q \rightarrow e^{i \alpha}
\bar q
\eeq
(in four component language, $q \rightarrow e^{i \alpha \gamma_5}
q$).
In particular, we can define a current:
\beq
j^{\mu}_5 = \bar q \gamma_5 \gamma_{\mu} q,
\eeq
and classically,
\beq
\partial_{\mu} j^{\mu}_5 = m \bar q \gamma_5 q.
\eeq
Under a transformation by an infinitesmal angle
$\alpha$ one would expect
\beq
{\delta L} = \alpha \partial_{\mu} j^{\mu}_5 = m \alpha \bar q
\gamma_5 q.
\eeq
But what we have just discovered is that the divergence of the
current contains another, $m$-independent, term:
\beq
\partial_{\mu} j^{\mu}_5 = m \bar q \gamma_5 q
+{1 \over 32 \pi^2}  F \tilde F. \eeq

This
anomaly can be derived in a number of other ways.  One can
define, for example, the current by ``point splitting," \beq
j^{\mu}_5 = \bar q(x+i\epsilon)e^{i\int_x^{x+\epsilon}dx^{\mu}A_{\mu}} q(x) \eeq
Because operators in quantum field theory are singular at short
distances, the Wilson line makes a finite contribution.  Expanding
the exponential carefully, one recovers the same expression for
the current.  A beautiful derivation, closely related to that we have
performed above, is due to Fujikawa, described in
\cite{anomalies}.  Here one considers the
anomaly as arising from a lack of invariance of the path integral
measure.  One carefully evaluates the Jacobian associated with the
change of variables $q \rightarrow q(1+i\gamma_5 \alpha)$, and
shows that it yields the same result\cite{anomalies}.
We will do a calculation along these lines in a two dimensional
model shortly.

The anomaly has important consequences in physics which will not
be the subject of the lecture today, but it is worth at least
listing a few before we proceed:
\begin{itemize}
\item
$\pi^o$ decay:
the divergence of
the axial
isospin current,
\beq
(j_5^3)^{\mu} = \bar u \gamma_5 \gamma^{\mu} \bar u - \bar d \gamma_5
\gamma^{\mu} d
\eeq
has an anomaly due to electromagnetism.  This gives rise to a
coupling of the $\pi^o$ to two photons, and the correct
computation of the lifetime was one of the early triumphs of the
theory of quarks with color.
\item
Anomalies in gauge currents signal an inconsistency in a theory.
They mean that the gauge invariance, which is crucial to the whole
structure of gauge theories (e.g. to the fact that they are
simultaneously unitary and lorentz invariant) is lost.
The absence of gauge anomalies is one of the striking
ingredients of the standard model, and it is also crucial in
extensions such as string theory.
\end{itemize}

Our focus in these lectures will be on another aspect of this anomaly:  the
appearance of an additional parameter
in the standard model, and the associated ``Strong CP
problem."

What we have just learned is that, if in our simple model above,
we require that the quark masses are real, we must allow for the
possible appearance in the lagrangian of the standard model, of the
$\theta$-terms of eqn. \ref{thetaterm}.  \footnote{In principle we must allow
a similar term, for the weak interactions.  However, $B+L$ is
a classical symmetry of the renormalizable interactions of the standard model.
This symmetry is anomalous, and can be used to remove the weak
$\theta$ term.  In the presence of higher dimension
$B+L$-violating terms, this is no longer true, but any effects of
$\theta$ will be extremely small, suppressed by $e^{-2
\pi/\alpha_w}$ as well as by powers of some large mass scale.}
 This term, however,
can be removed by a $B+L$ transformation.
What are the consequences of these terms?  We will focus on the
strong interactions, for which these terms are most important.
At first sight, one might guess that these terms are in fact of no
importance.  Consider, first, the case of QED.  Then
\beq
\int d^4 x F \tilde F
\eeq
is a the integral of a total divergence,
\beq
F \tilde F = \vec E \cdot \vec B = {1 \over 2}\partial_{\mu}
\epsilon^\mu_{\nu \rho \sigma} A^{\nu} F^{\rho \sigma}.
\eeq


As a result, this term does not
contribute to the classical equations of motion.  One might expect
that it does not contribute quantum mechanically either.  If we
think of the Euclidean path integral, configurations of finite
action have field strengths, $F_{\mu \nu}$ which fall off faster
than $1/r^2$ (where $r$ is the Euclidean distance), and $A$ which
falls off faster than $1/r$, so one can neglect surface terms in
${\cal L}_{\theta}$.

(A parenthetical remark:  This is almost correct.  However, if there are magnetic monopoles
there is a subtlety, first pointed out by
Witten\cite{witteneffect}.  Monopoles can carry electric charge.
In the presence of the $\theta$ term, there is an extra source for
the electric field at long distances, proportional to $\theta$ and
the monopole charge.  So the electric charges are given by:
\beq
Q = n_e e - {e \theta n_m \over 2 \pi}
\eeq
where $n_m$ is the monopole charge in units of the Dirac
quantum.)

In the case of non-Abelian gauge theories, the situation is more
subtle.  It is again true that $F \tilde F$ can be written as a
total divergence:
\beq
F \tilde F = \partial^{\mu} K_{\mu}~~~~~K_{\mu} = \epsilon_{\mu
\nu \rho \sigma}(A_{\nu}^{a} F_{\rho \sigma}^a
-{2 \over 3} f^{abc} A_{\nu}^a A_{\rho}^b A_{\sigma}^c).
\eeq
But now the statement that $F$ falls faster than $1/r^2$ does not
permit an equally strong statement about $A$.  We will see
shortly that there are finite action configurations --
finite action classical solutions -- where $F \sim {1 \over r^4}$,
but $A \rightarrow {1 \over
r}$, so that the surface term cannot be neglected.  These terms
are called instantons.  It is because of this that $\theta$ can
have real physical effects.

\section{A Two Dimensional Detour}

Before considering four dimensions with all of its complications,
it is helpful to consider two dimensions.  Two dimensions are
often a poor analog for four, but for some of the issues we are
facing here, the parallels are extremely close.  In these
two dimensional examples, the physics is more manageable, but
still rich.

\subsection{The Anomaly In Two Dimensions}

Consider, first, electrodynamics of a massless fermion in two
dimensions.  Let's investigate the anomaly.
The point-splitting method is particularly convenient here.  Just
as in four dimensions, we write:
\beq
j^{\mu}_5 = \bar \psi(x+i\epsilon)e^{i\int_x^{x+\epsilon}A_{\rho}dx^{\rho}}
\gamma^{\mu}  \psi(x)
\eeq
For very small $\epsilon$, we can pick up the leading singularity
in the product of $\psi(x+\epsilon) \psi$ by using the operator
product expansion, and noting that (using naive dimensional
analysis) the leading operator is the unit operator, with
coefficient proportional to $1/\epsilon$.  We can read
off this term by taking the vacuum expectation value, i.e. by
simply evaluating the propagator.  String theorists are particularly
familiar with this Green's function:
\beq
\langle \bar \psi(x+\epsilon) \psi(x) \rangle
={ 1 \over 2 \pi} {\epsilonslash \over \epsilon^2}
\eeq
Expanding the factor in the exponential to order $\epsilon$
gives
\beq
\partial_{\mu} j^{\mu}_5 = {\rm naive~ piece} +
{i \over 2 \pi}\partial_{\mu} \epsilon_{\rho} A^{\rho}
{\rm tr} { \epsilonslash \over \epsilon^2} \gamma^{\mu}
\gamma^5.
\eeq
Taking the trace gives $\epsilon_{\mu \nu} \epsilon^{\nu}$;
averaging $\epsilon$ over angles $(<\epsilon_{\mu} \epsilon_{\nu}>
= {1 \over 2} \eta_{\mu \nu} \epsilon^2)$, yields
\beq
\partial_{\mu}j^{\mu}_5 = {1 \over 4 \pi} \epsilon_{\mu \nu}F^{\mu
\nu}.
\eeq

\noindent
{\bf Exercise}:  Fill in the details of this computation, being
careful about signs and factors of $2$.

This is quite parallel to the situation in four dimensions.  The
divergence of the current is itself a total derivative:
\beq
\partial_{\mu} j^{\mu}_5 = {1 \over 2 \pi} \epsilon_{\mu \nu}
\partial^{\mu} A^{\nu}.
\eeq
So it appears possible to define a new current,
\beq
J^{\mu} = j^{\mu}_5 -{1 \over 2  \pi} \epsilon_{\mu \nu}
\partial^{\mu} A^{\nu}
\eeq
However, just as in the four dimensional case, this current is not
gauge invariant.  There is a familiar field
configuration for which $A$ does not fall off at infinity:  the
field of a point charge.  Indeed, if one has charges, $\pm \theta$ at
infinity, they give rise to a constant electric field,
$F_{oi} = e \theta$.  So $\theta$ has a very simple interpretation
in this theory.

It is easy to see that physics is periodic in $\theta$.
For $\theta >q$, it is energetically favorable to
produce a pair of charges from the vacuum which shield the charge
at $\infty$.

\subsection{The CP$^N$ Model:  An Asymptotically Free Theory}

The model we have considered so
far is not quite like QCD in at least two ways.  First, there
are no instantons; second, the coupling $e$ is dimensionful.  We
can obtain a theory closer to QCD
by considering the $CP^N$ model (our treatment here will follow
closely the treatment in Peskin and Schroeder's problem
13.3\cite{anomalies}).
This model starts
with a set of fields, $z_i$, $i=1,\dots N+1$.  These fields live
in the space $CP^N$.  This space is defined by the constraint:
\beq
\sum_i \vert z_i \vert^2 =1;
\eeq
in addition, the point $z_i$ is equivalent to $e^{i \alpha} z_i$.
To implement the first of these constraints,
we can add to the action a lagrange multiplier field,
$\lambda(x)$.  For the second,
we observe that the identification of points in the
``target space," $CP^N$, must hold at every point in {\it ordinary
space-time}, so this is a $U(1)$ gauge symmetry.   So
introducing a gauge field, $A_{\mu}$, and the corresponding
covariant derivative, we want to study the lagrangian:
\beq
{\cal L} = {1 \over g^2} [\vert D_{\mu} z_i \vert^2 - \lambda(x)(\vert
z_i \vert^2 -1)]
\eeq
Note that there is no kinetic term for $A_{\mu}$, so we can simply
eliminate it from the action using its equations of motion.  This
yields
\beq
{\cal L} = {1 \over g^2} [\vert \partial_{\mu} z_j \vert^2 +
\vert z_j^* \partial_{\mu} z_j \vert^2 ]
\eeq
It is easier to proceed, however, keeping $A_{\mu}$ in the action.
In this case, the action is quadratic in $z$, and we can integrate
out the $z$ fields:
\beq
Z= \int [dA] [d \lambda] [dz_j] exp[-{\cal L}] \\
= \int[dA] [d \lambda] e^{\int d^2 x \Gamma_{eff}[A,\lambda]}
\label{cpnintegral}
\eeq
$$
~~~~~~~=\int [dA] [d \lambda]exp[-N {\rm tr} \log(-D^2-\lambda) -{1
\over g^2} \int d^2 x \lambda]
$$

\subsection{The Large $N$ Limit}

By itself, the result of  eqn. \ref{cpnintegral}
is still rather complicated.  The fields $A_{\mu}$ and
$\lambda$ have complicated, non-local interactions.   Things
become much simpler if one takes the ``large $N$ limit", a limit
where one takes $N \rightarrow \infty$ with $g^2 N$ fixed.
In this case, the interactions of $\lambda$ and $A_{\mu}$ are
suppressed by powers of $N$.  For large $N$, the
path integral is dominated by a single field configuration, which
solves
\beq
{\delta \Gamma_{eff} \over \delta \lambda}=0
\eeq
or, setting the gauge field to zero,
\beq
N\int { d^2 k \over (2 \pi)^2}{1 \over k^2 + \lambda}= {1 \over
g^2},
\label{lambdaeqn}
\eeq
giving
\beq
\lambda = m^2 = M \exp[-{2 \pi \over g^2 N}].
\eeq
Here, $M$ is a cutoff required because the integral in eqn.
\ref{lambdaeqn} is divergent.  This result is remarkable.
One has
exhibited dimensional transmutation:  a theory which is
classically scale invariant contains non-trivial
masses, related in a renormalization-group invariant
fashion to the cutoff.
This is the phenomenon which in QCD
explains the masses of the proton, neutron, and other dimensionful
quantities.  So the theory is quite analogous to QCD.
We can read off the leading term in the $\beta$-function
from the familiar formula:
\beq
m = M e^{-\int {dg \over \beta(g)}}
\eeq
so, with
\beq
\beta(g) = -{1 \over 2 \pi} g^3 b_o
\eeq
we have $b_o = 1$.

But most important for our purposes, it is interesting to explore
the question of $\theta$-dependence.  Again, in this theory, we
could have introduced a $\theta$ term:
\beq
{\cal L}_{\theta} = {\theta \over 2 \pi} \int d^2 \epsilon_{\mu
\nu}F^{\mu \nu},
\eeq
where $F_{\mu \nu}$ can be expressed in terms of the fundamental
fields $z_j$.  As usual, this is the integral of
a total divergence.  But precisely as in the case of $1+1$
dimensional electrodynamics we discussed above, this term is
physically important.  In perturbation theory in the model, this
is not entirely obvious.  But using our reorganization of the
theory at large $N$, it is.  The lowest order action for $A_{\mu}$
is trivial, but at one loop (order $1/N$), one generates a kinetic
term for $A$ through the usual vacuum polarization loop:
\beq
{\cal L}_{kin}= {N \over 2 \pi m^2} F_{\mu \nu}^2.
\eeq
At this order, the effective theory consists of the gauge field,
then, with coupling $e^2 = {2 \pi m^2 \over N}$, and some coupling
to a dynamical, massive field $\lambda$.
As we have already argued, $\theta$ corresponds to a
non-zero background electric field due to charges at infinity,
and the theory clearly can have non-trivial $\theta$-dependence.

There is, in addition,
the possibility of including other light fields, for example
massless fermions.  In this case, one can again have an anomalous
$U(1)$ symmetry.
There is then no $\theta$-dependence, since it is possible to
shield any charge at infinity.  But there is non-trivial breaking
of the symmetry.  At low energies, one has now a theory with a fermion
coupled to a dynamical $U(1)$ gauge field.  The breaking
of the associated $U(1)$ in such a theory is a well-studied
phenomenon\cite{kogutsusskind}.

\noindent
{\bf Exercise:}  Complete Peskin and Schroeder, Problem 13.3.

\subsection{The Role of Instantons}

There is another way to think about the breaking of the $U(1)$
symmetry and $\theta$-dependence in this theory.  If one considers
the Euclidean functional integral, it is natural to look for
stationary points of the integration, i.e. for classical solutions
of the Euclidean equations of motion.  In order that they
be potentially important, it is necessary that these solutions
have finite action, which means that they must be localized in
Euclidean space and time.  For this reason, such solutions were dubbed
``instantons" by 't Hooft. Such solutions are not
difficult to find in the $CP^N$ model; we will
describe them briefly below.  These solutions carry
non-zero values of the topological charge,
\beq
{1 \over 2 \pi}\int d^2 x \epsilon_{\mu \nu}F_{\mu
\nu}=n\eeq
and have an action
$2 \pi n$.  As a result, they contribute to the
$\theta$-dependence; they give
a contribution to the functional integral:
\beq
Z_{inst} = e^{-2 \pi n\over g^2} e^{in \theta} \int [d \delta z_j] e^{-\delta
z_i
{\delta^2 S \over \delta z_i \delta z_j} \delta z_j} + \dots
\eeq
It follows
that:
\begin{itemize}
\item
Instantons generate $\theta$-dependence
\item
In the large $N$ limit, instanton effects are, formally, highly
suppressed, much smaller that the effects we found in the large
$N$ limit
\item
Somewhat distressingly, the functional integral above can not be systematically
evaluated.  The problem is that the classical theory is scale
invariant, as a result of which, instantons come in a variety of
sizes.  $\int [d \delta z]$ includes an integration over all
instanton sizes, which diverges for large size (i.e. in the
infrared).  This prevents a systematic evaluation of the effects
of instantons in this case.  At high
temperatures\cite{affleckcpn}, it is possible to do the
evaluation, and instanton effects are, indeed, systematically
small.
\end{itemize}

It is easy to construct the instanton solution in the case of
$CP^1$.  Rather than write the theory in terms of a gauge field,
as we have done above, it is convenient to parameterize the theory
in terms of a single complex field, $\phi$.  One can, for example,
define
$\phi = z_1/z_2$.  Then, with a bit of algebra, one can show that the
action for $\phi$ takes the form:
\beq
{\cal L} = (\partial_{\mu}\phi \partial_{\mu} \phi^*)
{1 \over 1 + \phi^* \phi} - {\phi^* \phi \over
(1 + \phi^* \phi)^2.}
\label{cpnaction}
\eeq
One can think of the field $\phi$ as living on the space with
metric given by the term in parenthesis, $g_{\phi \phi^*}.$  One can show that this
is the metric one obtains if one stereographically maps the sphere
onto the complex plane.
This mapping, which you may have seen in your math methods
courses, is just:
\beq
z = {x_1 + i x_2 \over 1-x_3};
\eeq
The inverse is
\beq
x_1 = {z + z^* \over 1 + \vert z \vert^2}
~~~~~x_2 = {z-z^* \over i (1 + \vert z \vert^2)}
\label{stereographic}
\eeq
$$~~~~~~~x_3 = {\vert z \vert^2 -1 \over \vert z \vert^2 + 1}.
$$

It is straightforward to write down the equations of motion:
\beq
\partial^2 \phi g_{\phi \bar \phi} + \partial_{\mu}\phi
(\partial_{\mu} \bar \phi {\partial g \over \partial \bar \phi}
+ \partial_{\mu} \phi{\partial g \over \partial \phi}) = 0
\label{solution}
\eeq
Now calling the space time coordinates $z=x_1 + i x_2$,
$z^* = x_1-ix_2$, you can see that if $\phi$ is analytic,
the equations of motion are satisfied!  So a simple solution,
which you can check has finite action, is
\beq
\phi(z) = \rho z.
\label{instanton}
\eeq
In addition to evaluating the action, you can evaluate the
topological charge,
\beq
{1 \over 2 \pi} \int d^2 x\epsilon_{\mu \nu} F^{\mu \nu} =1
\label{charge}
\eeq
for this solution.  More generally, the topological charge measures the
number of times that $\phi$ maps the complex plane into the
complex plane; for $\phi =z^n$, for example, one has charge $n$.

\noindent
{\bf Exercise:}
Verify that the action of eqn. \ref{cpnaction} is equal to
\beq
{\cal L} = g_{\phi, \phi^*} \partial_{\mu} \phi \partial_{\mu} \phi^*
\eeq
where $g$ is the metric of the sphere in complex coordinates, i.e.
it is the line element $dx_1^2 + dx_2^2 + dx_3^2$
expressed as $g_{z, z}dz ~dz + g_{z,z^*} dz~dz^* + g_{z^* z} dz^*
~dz + g_{dz^* dz^*} dz^* dz^*$.  A model with an action of this
form is called a ``Non-linear Sigma Model;"  the idea is that the
fields live on some ``target" space, with metric $g$.  Verify
eqns. \ref{cpnaction},\ref{stereographic}.


More generally, $\phi = {az + b \over c z + d}$ is a solution with
action $2\pi$.  The parameters $a,\dots d$ are called collective
coordinates.  They correspond to the symmetries of translations,
dilations, and rotations, and special conformal transformations
(forming the group $SL(2,C)$).
In other words, any given finite action solution breaks the
symmetries.  In the path integral, the symmetry of Green's
functions is recovered when one integrates over the collective
coordinates.  For translations, this is particularly simple.
If one studies a Green's function,
\beq
<\phi(x) \phi(y)> \approx \int d^2 x_o  \phi_{cl}(x-x_o)
\phi_{cl}(y-y_o) e^{-S_o}
\eeq
The precise measure is obtained by the Fadeev-Popov method.
Similarly, the integration over the parameter $\rho$ yields
a factor
\beq
\int d \rho \rho^{-1} e^{-{2\pi \over g^2(\rho)}} \dots
\eeq
Here the first factor follows on dimensional grounds.  The second
follows from renormalization-group considerations.  It can be
found by explicit evaluation of the functional determinant\cite{thooft}.
Note that, because of asymptotic freedom, this means that typical
Green's functions will be divergent in the infrared.


  There are many other features of this instanton one can
consider.  For example, one can consider adding massless fermions
to the model, by simply coupling them in a gauge-invariant way
to $A_{\mu}$.  The resulting theory has a chiral $U(1)$ symmetry,
which is anomalous.  In the presence of an instanton, one
can easily construct normalizable fermion zero modes (the Dirac
equation just becomes the statement that $\psi$ is analytic).
As a result, Green's functions computed in the
instanton background do not respect the axial $U(1)$ symmetry.
But rather than get too carried away with this model
(I urge you to get a little carried away and play with it a bit),
let's proceed to four dimensions, where we will see very similar
phenomena.

\section{Real QCD}

The model of the previous section mimics many features of real
QCD.  Indeed, we will see that much of our discussion can be
carried over, almost word for word, to the observed strong
interactions.  This analogy is helpful, given that in QCD we have
no approximation which gives us control over the theory comparable
to that which we found in the large $N$ limit of the $CP^N$ model.
As in that theory:
\begin{itemize}
\item
There is a $\theta$ parameter, which appears as an integral over
the divergence of a non-gauge invariant current.
\item
There are instantons, which indicate that there should be real
$\theta$-dependence.  However, instanton effects cannot be
considered in a controlled approximation, and there is no clear
sense in which $\theta$-dependence can be understood as arising
from instantons.
\item
There is another approach to the theory, which shows that the
$\theta$-dependence is real, and allows computation of these
effects.  In QCD, this is related to the breaking of chiral
symmetries.
\end{itemize}

\subsection{The Theory and its Symmetries}

While it is not in the spirit of much of this school, which is
devoted to the physics of heavy quarks, it is sufficient, to
understand the effects of $\theta$, to focus on only the light
quark sector of QCD.  For simplicity in writing some of the
formulas, we will consider two light quarks; it is not difficult
to generalize the resulting analysis to the case of three.
It is believed that the masses of the $u$ and $d$ quarks are of
order $5$ MeV and $10$ MeV, respectively, much lighter than the
scale of QCD.  So we first consider an idealization of the theory
in which these masses are set to zero.  In this limit, the theory
has a symmetry $SU(2)_L \times SU(2)_R$.  This symmetry is
spontaneously broken to a vector $SU(2)$.  The three resulting
Goldstone bosons are the $\pi$ mesons.  Calling
\beq
q = \left ( \matrix{u \cr d} \right ) ~~~~\bar q = \left ( \matrix{ \bar u \cr
\bar d} \right ),
\eeq
the two $SU(2)$ symmetries act separately on $q$ and $\bar q$
(thought of as left handed fermions).  The order parameter for the
symmetry breaking is believed to be the condensate:
\beq
{\cal M}_o = \langle \bar q q \rangle.
\eeq
This indeed breaks the symmetry down to the vector sum.  The
associated
Goldstone bosons are the $\pi$ mesons. One can think of the
Goldstone bosons as being associated with a slow variation of the
expectation value in space, so we can introduce a composite operator
\beq
{\cal M} =  \bar q q = M_o e^{i{ \pi_a(x) \tau_a \over f_{\pi}}}\left (
\matrix{1 & 0 \cr 0 & 1} \right )
\eeq
The quark mass term in the lagrangian is then (for simplicity
writing $m_u=m_d = m_q$; more generally one should introduce a
matrix)
\beq
m_q {\cal M}.
\eeq
Expanding $\cal M$ in powers of $\pi/f_{\pi}$, it is clear that
the minimum of the potential occurs for $\pi_a =0$.  Expanding to
second order, one has
\beq
m_{\pi}^2 f_\pi^2 = m_q M_o.
\eeq

But we have been a bit cavalier about the symmetries.  The theory
also has two $U(1)$'s;
\beq
q \rightarrow e^{i \alpha }q ~~~~~\bar q \rightarrow e^{i \alpha}
\bar q
\\
q \rightarrow e^{i \alpha }q ~~~~~\bar q \rightarrow e^{-i \alpha}
\bar q
\eeq
The first of these
is baryon number and it is not chiral (and is not broken
by the condensate).  The second is the axial
$U(1)_5$; It is also broken by the condensate.
So, in addition to the pions, there should be another approximate
Goldstone boson.  The best candidate is the $\eta$, but, as we
will see below (and as you will see further in Thomas's lectures),
the $\eta$ is too heavy to be interpreted in this way.
The absence of this fourth (or in the case of three light quarks,
ninth) Goldstone boson is called the $U(1)$ problem.

The $U(1)_5$ symmetry suffers from an anomaly,
however, and we might hope that this has something to do with the
absence of a corresponding Goldstone boson.  The anomaly is given
by
\beq
\partial_{\mu} j^{\mu}_5 = {2 \over 32 \pi^2} F \tilde F
\eeq
Again, we can write the right hand side as a total divergence,
\beq
F \tilde F = \partial_{\mu} K^{\mu}
\eeq
where
\beq
K_{\mu} = \epsilon_{\mu
\nu \rho \sigma}(A_{\nu}^{a} F_{\rho \sigma}^a
-{2 \over 3} f^{abc} A_{\nu}^a A_{\rho}^b A_{\sigma}^c).
\eeq
So if it is true that this term accounts for the absence of the
Goldstone boson, we need to show that there are important
configurations in the functional integral for which the rhs does
not vanish rapidly at infinity.

\subsection{Instantons}

It is easiest to study the Euclidean version of the theory.
This is useful if we are interested in very low energy processes,
which can be described by an effective action expanded about zero
momentum.  In the functional integral,
\beq
Z= \int [dA] [dq][d\bar q] e^{-S}
\eeq
it is natural to look for stationary points of the effective
action, i.e. finite action, classical solutions of the theory in imaginary time.
The Yang-Mills equations are complicated, non-linear equations,
but it turns out that, much as in the
$CP^N$ model, the instanton solutions can be
found rather easily\cite{polyakov}.  The following tricks simplify
the construction, and turn out to yield the general solution.
First, note that the
Yang-Mills action satisfies an inequality:
\beq
\int(F \pm \tilde F)^2 = \int (F^2 + \tilde F^2 \pm
2 F \tilde F) = \int (2 F^2 + 2F \tilde F) \ge 0.
\eeq
So the action is bounded  $\int F \tilde F$, with the bound
being saturated when
\beq
F = \pm \tilde F
\label{selfduality}
\eeq
i.e. if the gauge field is (anti) self dual.\footnote{This
is not an accident, nor was the analyticity condition
in the $CP^N$ case.  In both cases, we can add fermions so that the
model is supersymmetric.  Then one can show that if some of the
supersymmetry generators, $Q_{\alpha}$ annihilate a field
configuration, then the configuration is a solution.  This
is a first order condition; in the Yang-Mills case, it implies
self-duality, and in the $CP^N$ case it requires analyticity.}
This equation is a
first order equation, and it is easy to solve if one first
restricts to an $SU(2)$ subgroup of the full gauge
group. One makes the
ansatz that the solution should be invariant under a combination
of ordinary rotations and global $SU(2)$ gauge transformations:
\beq
A_{\mu} = f(r^2)  + h(r^2) \vec x \cdot \vec \tau
\eeq
where we are using a matrix notation for the gauge fields.
One can actually make a better guess:
define the gauge transformation:
\beq
g(x) = {x_4 + i \vec x \cdot \vec \tau \over r}
\eeq
and take
\beq
A_{\mu} = f(r^2) g \partial_{\mu} g^{-1}
\label{yminstanton}
\eeq
Then plugging in the Yang-Mills equations yields:
\beq
f= {r^2 \over r^2 + \rho^2}
\label{formoff}
\eeq
where $\rho$ is an arbitrary quantity with dimensions of length.
The choice of origin here is arbitrary; this can be remedied
by simply replacing $x \rightarrow x-x_o$ everywhere in these
expressions, where $x_o$ represents the location of the instanton.

\noindent
{\bf Exercise}  Check that eqns. \ref{yminstanton},\ref{formoff}
solve \ref{selfduality}.

>From this solution, it is clear why $\int \partial_{\mu} K^{\mu}$ does not vanish for
the solution:  while $A$ is a pure gauge at infinity, it falls
only as $1/r$.  Indeed, since $F = \tilde F$, for this solution,
\beq
\int F^2 = \int \tilde F^2 = 32 \pi^2
\eeq
This result can also be understood topologically.  $g$ defines a
mapping from the ``sphere at infinity" into the gauge group.  It
is straightforward to show that
\beq
{1 \over 32 \pi^2}\int d^4 x F \tilde F
\eeq
counts the number of times $g$ maps the sphere at infinity into
the group (one for this specific example; $n$ more generally).
We do not have time to explore all of this in detail; I urge you
to look at Sidney Coleman's lecture, ``The Uses of
Instantons"\cite{coleman}.  To actually do calculations, 't Hooft
developed some notations which are often more efficient than those
described above\cite{thooft}.

So we have exhibited potentially important contributions to the
path integral which violate the $U(1)$ symmetry.  How does this
violation of the symmetry show up?  Let's think about the path
integral in a bit more detail.  Having found a classical solution,
we want to integrate about small fluctuations about it:
\beq
Z= e^{-{8 \pi^2 \over g^2}} e^{i\theta}\int [d\delta A] [dq][d\bar q]
e^{i\delta^2 S}
\eeq
Now $S$ contains an explicit factor of $1/g^2$.  As a result, the
fluctuations are formally suppressed by $g^2$ relative to the leading
contribution.  The
one loop functional integral yields a product of determinants
for the fermions, and of inverse square root determinants
for the bosons.  Consider, first, the integral over the fermions.
It is straightforward, if challenging, to evaluate the
determinants\cite{thooft}.  But if the quark masses are zero, the
fermion functional integrals are zero, because there is a zero
mode for each of the fermions, i.e. for both $q$ and $\bar q$
there is a normalizable solution of the equation:
\beq
\Dslash u = 0~~~~\Dslash \bar u =0
\eeq
and similarly for $d$ and $\bar d$.
It is straightforward to construct these solutions:
\beq
u = {\rho \over (\rho^2 + (x-x_o)^2)^{3/2}} \zeta
\eeq
where $\zeta$ is a constant spinor,
and similarly for $\bar u$, etc.

This means that in order for the path integral to be
non-vanishing, we need to include insertions of enough $q$'s and
$\bar q$'s to soak up all of the zero modes.
In other words, non-vanishing Green's functions have the form
\beq
\langle \bar u u \bar d d \rangle
\eeq
and violate the symmetry.  Note that the symmetry violation is
just as predicted from the anomaly equation:
\beq
\Delta Q_5 = 4 {1 \over  \pi^2} \int d^4 x F \tilde F =4
\eeq

However, the calculation we have described here is not self
consistent.  The difficulty is that among the variations of the
fields we need to integrate over are changes in the location of
the instanton (translations), rotations of the instanton, and
scale transformations.  The translations are easy to deal with;
one has simply to integrate over $x_o$ (one must also include a
suitable Jacobian factor\cite{coleman}).  Similarly, one must
integrate over $\rho$.  There is a power of $\rho$ arising from
the Jacobian, which can be determined on dimensional grounds.
For our Green's function above, for example, which has dimension
$6$, we have (if all of the fields are evaluated at the same
point),
\beq
\int d\rho \rho^{-7}.
\eeq
However, there is additional $\rho$-dependence because the quantum
theory violates the scale symmetry.  This can be understood by
replacing
$g^2 \rightarrow g^2(\rho)$ in the functional integral, and using
\beq
e^{-{8\pi^2 g^2(\rho)}}
\approx (\rho M)^{b_o}
\eeq
for small $\rho$.  For $3$ flavor QCD, for example, $b_o = 9$, and
the $\rho$ integral diverges for large $\rho$.  This is just the
statement that the integral is dominated by  the infrared, where
the QCD coupling becomes strong.

So we have provided some evidence that the $U(1)$ problem is
solved in QCD, but no reliable calculation.  What about
$\theta$-dependence?  Let us ask first about $\theta$-dependence
of the vacuum energy.  In order to get a non-zero result, we need
to allow that the quarks are massive.  Treating the mass as a
perturbation, we obtain
\beq
E(\theta) = C \Lambda_{QCD}^9 m_u m_d \cos(\theta) \int d\rho
\rho^{-3} \rho^9.
\eeq
So again, we have evidence for $\theta$-dependence, but cannot do
a reliable calculation.  That we cannot do a calculation
should not be a surprise.  There is
no small parameter in QCD to use as an expansion parameter.
Fortunately, we can use other facts which we know about the strong
interactions to get a better handle on both the $U(1)$ problem and
the question of $\theta$-dependence.

Before continuing, however, let us consider the weak interactions.
Here there is a small parameter, and there are no infrared
difficulties, so we might expect instanton effects to be small.
The analog of the $U(1)_5$ symmetry in this case is baryon number.
Baryon number has an anomaly in the standard model, since all of the quark
doublets have the same sign of the baryon number.  't Hooft
realized that one could actually use instantons, in this case, to
compute the violation of baryon number.  Technically, there are
no finite action Euclidean solutions in this theory; this follows,
as we will see in a moment, from a simple scaling argument.
However, 't Hooft realized that one can construct important
configurations of non-zero topological charge by starting with the
instantons of the pure gauge theory and perturbing them.  If one
simply takes such an instanton, and plugs it into the action, one
necessarily finds a correction to the action of the form
\beq
\delta S = {1 \over g^2} v^2 \rho^2.
\eeq
This damps the $\rho$ integral at large $\rho$, and leads to a
convergent result.  Affleck showed how to develop this into a
systematic computation\cite{affleck}.  Note that from this, one can see that
baryon number violation occurs in the standard model, and that the
rate is incredibly small,
proportional to $e^{-2{\pi \over \alpha_W}}$.

\subsection{Real QCD and the U(1) Problem}

In real QCD, it is difficult to do a reliable calculation which shows
that there is not an extra Goldstone boson, but the instanton
analysis we have described makes clear that there is no reason to
expect one.  Actually, while perturbative and semiclassical
(instanton) techniques have no reason to give reliable results,
there are two approximation methods techniques which are
available.
The first is large $N$, where one now allows the $N$ of $SU(N)$ to
be large, with $g^2 N$ fixed.  In contrast to the case of $CP^N$,
this does not permit enough simplification to do explicit
computations, but it does allow one to make qualitative statements
about the theory.
available in QCD.   Witten
has pointed out a way in which one can at relate the
mass of the $\eta$ (or $\eta^\prime$ if one is thinking in terms
of $SU(3) \times SU(3)$ current algebra) to quantities in a theory
without quarks.  The point is to note that the anomaly is an
effect suppressed by a power of $N$, in the large $N$ limit.
This is because the loop diagram contains a
factor of $g^2$ but not of $N$.  So, in large $N$, it can be treated
as a perturbation, and the the $\eta$ is massless.
$\partial_{\mu} j^{\mu}_5$ is like a creation operator for the $\eta$, so
(just like $\partial_{\mu}j^{\mu ~3}_5$ is a creation operator for
the $\pi$ meson), so one can compute the mass if one knows the
correlation function, at zero momentum, of
\beq
\langle \partial_{\mu} j^{\mu}_5(x) \partial_{\mu} j^{\mu}_5(y) \rangle
\propto {1\over N^2} \langle F(x) \tilde F(x) F(y) \tilde F(y) \rangle
\eeq
To leading order in the $1/N$ expansion, this correlation function
can be computed in the theory without quarks.
Witten argued that while this vanishes order by order in
perturbation theory, there is no reason that this correlation
function need vanish in the full theory.  Attempts have been made
to compute this quantity both in lattice gauge theory and using
the AdS-CFT correspondence recently discovered in string theory.
Both methods give promising results.

So the U(1) problem should be viewed as solved, in the sense that
absent any argument to the contrary, there is no reason to think
that there should be an extra Goldstone boson in QCD.

The second approximation scheme which gives some control
of QCD is known as chiral perturbation theory.  The masses of the
$u$, $d$ and $s$ quarks are small compared to the QCD scale, and
the mass terms for these quarks in the lagrangian can be treated
as perturbations.  This will figure in our discussion in the next
section.

\subsection{Other Uses of Instantons:  A Survey}

In the early days of QCD, it was hoped that instantons, being a
reasonably well understood non-perturbative effect,
 might give
insight into many aspects of the strong interactions.   Because of
the infrared divergences discussed earlier, this program proved to
be a disappointment.  There was simply no well-controlled
approximation to QCD in which instantons were important.  Indeed,
Witten\cite{wittencpn} stressed the successes of the large $N$
limit in understanding the strong interactions, and argued that in
this limit, anomalies could be important but instantons would be
suppressed exponentially.  This reasoning (which I urge you
to read) underlay much of our earlier discussion, which borrowed
heavily on this work.

In the years since Coleman's ``Uses of Instantons" was
published\cite{coleman}, many uses of instantons in controlled
approximations have been found.  What follows is an incomplete
list; I hope this will inspire some of you to read Coleman's
lectures and develop a deeper understanding of the subject.

\begin{itemize}
\item
$\theta$-dependence at finite temperature:
Within QCD, instanton calculations are reliable at high
temperatures.  So, for example, one can calculate the
$\theta$-dependence of the vacuum energy in the early universe,
and other quantities to which instantons give the leading
contribution\cite{gpy}.
\item
Baryon number violation in the standard model:  We have remarked
that this can be reliably calculated, though it is extremely
small.   However, as explained in \cite{coleman}, instanton
effects are associated with tunneling, and in the standard model,
they describe tunneling between states with different baryon
number.  It is reasonable to expect that baryon number violation
is enhanced at high temperature, where one has plenty of energy
to pass over the barrier without tunneling. This is indeed the case.  This baryon
number violation might be responsible for the matter-antimatter
asymmetry which we observe\cite{ckn}.
\item
Instanton effects in supersymmetric theories:  this has turned out
to be a rich topic.  Instantons, in many instances, are the
leading effects which violate non-renormalization theorems in
perturbation theory, and they can give rise to superpotentials,
supersymmetry breaking, and other phenomena.  More generally, they
have provided insight into a whole range of field theory and
string theory phenomena.
\end{itemize}

\section{The Strong CP Problem}

\subsection{$\theta$-dependence of the Vacuum Energy}

The fact that the anomaly resolves the $U(1)$ problem in QCD,
however, raises another issue.  Given that $\int d^4 x F \tilde F$
has physical effects, the theta term in the action has physical
effects as well.  Since this term is CP odd, this means that there
is the potential for strong CP violating effects.  These effects
should vanish in the limit of zero quark masses, since in this
case, by a field redefinition, we can remove $\theta$ from the
lagrangian.  In the presence of quark masses, the
$\theta$-dependence of many quantities can be computed.  Consider,
for example, the vacuum energy.
In QCD, the quark mass term in the lagrangian has the form:
\beq
{\cal L}_m = m_u \bar u u + m_d \bar d d + {\rm h.c.}
\eeq

Were it not for the anomaly, we could, by redefining the quark
fields, take $m_u$ and $m_d$ to be real.  Instead, we can define
these fields so that there is no $\theta F \tilde F$ term in the
action, but there is a phase in $m_u$ and $m_d$.  Clearly, we have
some freedom in making this choice.  In the case that $m_u$ and
$m_d$ are equal, it is natural to choose these phases to be the
same.  We will explain shortly how one proceeds when the masses
are different (as they are in nature).  We can, by convention,
take $\theta$ to be the phase of the overall lagrangian:
\beq
{\cal L}_m = (m_u \bar u u + m_d \bar d d) \cos(\theta/2) + {\rm h.c.}
\label{thetal}
\eeq

Now we want to treat this term as a perturbation.  At first order,
it makes a contribution to the ground state energy proportional to
its expectation value.
We have already argued that the quark bilinears have non-zero vacuum
expectation values, so
\beq
E(\theta) = (m_u + m_d) e^{i\theta} \langle \bar q q \rangle.
\label{vtheta}
\eeq

While, without a difficult non-perturbative calculation, we can't
calculate the separate quantities on the right hand side of this
expression, we can, using current algebra, relate them to measured
quantities.  A simple way to do this is to use the effective
lagrangian method (which will be described in more detail in
Thomas's lectures).  The basic idea is that at low energies, the
only degrees of freedom which can readily be excited in QCD are
the pions.  So parameterize $\bar q q$ as
\beq
\bar q q = \Sigma = <\bar q q> e^{i {\pi^a(x) \sigma^a \over 2 f_\pi}}
\eeq
We can then write the quark mass term as
\beq
{\cal L}_m =e^{i \theta} Tr M_q \Sigma.
\eeq
Ignoring the $\theta$ term at first, we can see, plugging in the
explicit form for $\Sigma$, that
\beq
m_{\pi}^2 f_{\pi}^2= (m_u+m_d) <\bar q q>.
\eeq
So the vacuum energy, as a function of $\theta$, is:
\beq
E(\theta) = m_{\pi}^2 f_{\pi}^2 \cos(\theta).
\eeq
This expression can readily be generalized to the case of three
light quarks by similar methods.  In any case, we now see that
there is real physics in $\theta$, even if we don't understand how
to do an instanton calculation.  In the next section, we will
calculate a more physically interesting quantity:  the neutron
electric dipole moment as a function of $\theta$.

\subsection{The Neutron Electric Dipole Moment}

As Scott Thomas will explain in much greater detail in his
lectures, the most interesting physical quantities to study in
connection with CP violation are electric dipole moments,
particularly that of the
neutron, $d_n$.  It has been possible to set
strong experimental limits on this quantity.  Using current
algebra, the leading contribution to the neutron electric dipole
moment due to $\theta$ can be calculated, and one obtains a limit $\theta <
10^{-9}$\cite{crewther}.  The original paper on the subject is quite readable.
Here we outline the main steps in the calculation; I urge you to
work out the details following the reference.
We will simplify the analysis by working in an exact
$SU(2)$-symmetric limit, i.e. by taking $m_u=m_d =m$.
We again treat the lagrangian of [\ref{thetal}] as a perturbation.
We can also understand how this term depends on the $\pi$ fields
by making an axial $SU(2)$ transformation on the quark fields.
In other words, a background $\pi$ field can be thought of as a
small chiral transformation from the vacuum.
Then, e.g., for the $\tau_3$ direction, $q \rightarrow (1+ i
\pi_3 \tau_3) q$ (the $\pi$ field parameterizes the
transformation), so the action becomes:
\beq
{m \over f_{\pi}} \pi_3( \bar q \gamma_5 q +
\theta \bar q  q)
\eeq
In other words, we have calculated a CP violating coupling of the
mesons to the pions.

\begin{figure}[htbp]
\centering
\centerline{\psfig{file=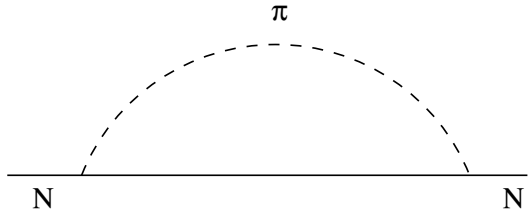,angle=-90,width=7cm}}
\caption{Diagram in which CP-violating coupling of the pion
contributes to $d_n$.}
\label{dngraph}
\end{figure}

This coupling is difficult to measure
directly, but it was observed in \cite{crewther} that this
coupling gives rise, in a calculable fashion, to a neutron
electric dipole moment.  Consider the graph of fig. \ref{dngraph}.  This
graph generates a neutron electric dipole moment,
if we take one coupling to be the standard pion-nucleon coupling,
and the second the coupling we have computed above.  The resulting
Feynman graph is infrared divergent; we cut this off at $m_{\pi}$,
while cutting off the integral in the ultraviolet at the QCD
scale.  Because of this infrared sensitivity, the low energy
calculation is reliable.  The exact result is:
\beq
d_n = g_{\pi N N}{- \theta m_u m_d \over f_{\pi} (m_u+m_d)}
\langle N_f \vert \bar q \tau^a q \vert N_i \rangle
\ln(M_N/m_{\pi}) {1 \over 4 \pi^2}M_N.
\eeq
The matrix element can be estimated using ordinary $SU(3)$,
yielding
$d_n = 5.2 \times 10^{-16} \theta ~{\rm cm}$.
The experimental bound gives $\theta < 10^{-9-10}.$
Understanding why CP violation is so small in the strong
interactions is the ``strong CP problem."

\section{Possible Solutions}

What should our attitude towards this problem be?  We
might argue that, after all, some Yukawa couplings are as small as
$10^{-5}$, so why is $10^{-9}$ so bad.  On the other hand,
we suspect that the smallness of the Yukawa couplings
is related to approximate symmetries, and that these
Yukawa couplings are telling us something.  Perhaps there is some
explanation of the smallness of $\theta$, and perhaps this is a
clue to new physics.
In this section we review some of the solutions which have been
proposed to understand the smallness of $\theta$.

\subsection{Massless u Quark}

Suppose the bare mass of the u quark was zero (i.e. at some high
scale, the $u$ quark mass were zero).  Then, by a redefinition of
the $u$ quark field, we could eliminate $\theta$ from the
lagrangian.  Moreover, as we integrated out physics from this high
scale to a lower scale, instanton effects would generate a small
$u$ quark mass.  In fact, a crude estimate suggests that this mass
will be comparable to the estimates usually made from current
algebra.  Suppose that we construct a Wilsonian action at a scale,
say, of order twice the QCD scale.  Call this scale $\Lambda_o$.
Then we would expect, on dimensional grounds, that the $u$ quark
mass would be of order:
\beq
m_u = {m_d m_s \over \Lambda_o}
\eeq
Now everything depends on what you take $\Lambda_o$ to be, and there
is much learned discussion about this.  The general belief seems
to be that the coefficient of this expression needs to be of order
three to explain the known facts of the hadron spectrum.  There is
contentious debate about how plausible this possibility is.

Note, even if one does accept this possibility, one would still
like to understand why the $u$ quark mass at the high scale is exactly zero (or
extremely small).  It is interesting that in string theory, one
knows of discrete symmetries which are anomalous, i.e. one has a
fundamental theory where there are discrete symmetries which can
be broken by very tiny effects.  Perhaps this could be the
resolution of the strong CP problem?

\subsubsection{CP as a Spontaneously Broken Symmetry}

A second possible solution comes from the observation that if the
underlying theory is $CP$ conserving, a ``bare" $\theta$ parameter
is forbidden.  In such a theory, the observed CP violation must
arise spontaneously, and the challenge is to understand why this
spontaneous CP violation does not lead to a $\theta$ parameter.
For example, if the low energy theory contains just the standard
model fields, then some high energy breaking of CP must generate
the standard model CP violating phase.  This must not generate
a phase in $\det m_q$, which would be a $\theta$ parameter.
Various schemes have been devised to accomplish this\cite{precursors,barrnelson}.
Without supersymmetry, they are generally invoked in the context of
grand unification.  There, it is easy to arrange that
the $\theta$ parameter vanishes at the tree level, including
only renormalizable operators.  It is then necessary to understand suppression
of loop effects and of the contributions
of higher dimension operators.  In the
context of supersymmetry, it turns out that understanding the
smallness of $\theta$ in such a framework, requires that the
squark mass matrix have certain special properties (there must
be a high degree of squark degeneracy, and the left right terms
in the squark mass matrices must be nearly proportional
to the quark mass matrix.)\cite{dkl}.

Again, it is interesting that string theory is a theory in which
CP is a fundamental (gauge) symmetry; its breaking is necessarily
spontaneous.  Some simple string models possess some of the
ingredients required to implement the ideas of \cite{precursors,barrnelson}.

\subsubsection{The Axion}

Perhaps the most popular explanation
of the smallness of $\theta$ involves a hypothetical particle
called the axion.  We present here a slightly updated version of
the original idea of Peccei and Quinn\cite{pq}.

Consider the vacuum energy as a function of $\theta$, eqn.
[\ref{vtheta}].  This energy has a minimum at $\theta=0$, i.e. at
the CP conserving point.  As Weinberg noted long ago, this is
almost automatic:  points of higher symmetry are necessarily
stationary points.  As it stands, this observation is not
particularly useful, since $\theta$ is a parameter, not a
dynamical variable.  But suppose that one has a particle, $a$,
with coupling to QCD:
\beq
{\cal L}_{axion} = (\partial_{\mu} a)^2 + {(a/f_a+ \theta) \over 32 \pi^2} F \tilde F
\eeq
$f_a$ is known as the axion decay constant.
Suppose that the rest of the theory possesses a symmetry,
called the Peccei-Quinn symmetry,
\beq
a \rightarrow a + \alpha
\eeq
for constant $\alpha$.  Then by a shift in $a$, one can eliminate
$\theta$.  $E(\theta$ is now $V(a/f_a)$, the potential energy of
the axion.  It has a minimum at $\theta=0$.  The strong CP problem
is solved.

One can estimate the axion mass by simply examining $E(\theta)$.
\beq
m_a^2 \approx {m_{\pi}^2 f_{\pi^2} \over f_a^2}.
\eeq
If $f_a \sim TeV$, this yields a mass of order KeV.  If $f_a \sim
10^{16}$ GeV, this gives a mass of order $10^{-9}$ eV.  As for the
$\theta$-dependence of the vacuum energy, it is not difficult to
get the factors straight using current algebra methods.  A
collection of formulae, with great care about factors of $2$,
appears in \cite{srednicki}

Actually, there are several questions one can raise about this
proposal:
\begin{itemize}
\item
Should the axion already have been observed?
In fact, as Scott Thomas
will explain in greater detail in his lectures, the couplings of
the axion to matter can be worked out in a straightforward way,
using the methods of current algebra (in particular of non-linear
lagrangians).  All of the couplings of the axion are suppressed by
powers of $f_a$.  So if $f_a$ is large enough, the axion is
difficult to see.  The strongest limit turns out to come from red
giant stars.  The production of axions is ``semiweak," i.e. it
only is suppressed by one power of $f_a$, rather than two powers
of $m_W$; as a result, axion emission is competitive with neutrino
emission until $f_a > 10^{10}$ GeV or so.
\item
As we will describe in a bit more detail below, the axion can be
copiously produced in the early universe.  As a result, there is
an {\it upper bound} on the axion decay constant.  In this case,
as we will explain below, the axion could  constitute the dark
matter.
\item
Can one search for the axion experimentally\cite{sikivie}?  Typically, the axion
couples not only to the $F \tilde F$ of QCD, but also to the same
object in QED.  This means that in a strong magnetic field, an
axion can convert to a photon.   Precisely this effect is being
searched for by a group at Livermore (the collaboration contains
members from MIT, University of Florida) and Kyoto.
The basic idea is to suppose that the dark matter in the halo
consists principally of axions.   Using a (superconducting) resonant cavity
with a high Q value in a large magnetic field, one searches
for the conversion of these axions into excitations
of the cavity.  The experiments
have already reached a level where they set interesting limits;
the next generation of experiments will cut a significant swath
in the presently allowed parameter space.
\item
The coupling of the axion to $F \tilde F$ violates the shift
symmetry; this is why the axion can develop a potential.  But this
seems rather paradoxical:  one postulates a symmetry, preserved
to some high degree of approximation, but which is not a symmetry;
it is at least broken by tiny QCD
effects.  Is this
reasonable?  To understand the nature of the problem,
consider one of the ways an axion
can arise.  In some approximation, we can suppose we have a global
symmetry under which a scalar field, $\phi$, transforms as
$\phi \rightarrow e^{i \alpha} \phi$.  Suppose, further, that
$\phi$ has an expectation value, with an associated
(pseudo)-Goldstone boson, $a$.  We can parameterize $\phi$ as:
\beq
\phi = f_a e^{ia/f_a}~~~~~\vert \langle    \phi \rangle \vert =
f_a
\eeq
If this field couples to fermions, so that they gain mass from its
expectation value, then at one loop we generate a coupling
$a F \tilde F$ from integrating out the fermions.  This calculation
is identical to the corresponding calculation for pions
we discussed earlier.  But we usually assume that global symmetries
in nature are accidents.  For example, baryon number is conserved in
the standard model simply because there are no gauge-invariant,
renormalizable operators which violate the symmetry.
We believe it is violated by higher dimension terms.
The global symmetry we postulate here is presumably an accident
of the same sort.  But for the axion, the symmetry must be
{\it extremely} good. For example, suppose
one has a symmetry breaking operator
\beq
\phi^{n+4} \over M_p^n
\eeq
Such a term gives a linear contribution to the axion potential of order
$f_a^{n+3} \over M_p^n$.  If $f_a \sim 10^{11}$, this swamps the would-be QCD
contribution ($m_{\pi}^2 f_{\pi}^2 \over f_a$) unless $n >
12$\cite{kamionkowski}!
\end{itemize}

This last objection finds an answer in string theory.  In this
theory, there are axions, with just the right properties, i.e.
there are symmetries in the theory which are {\it exact} in
perturbation theory, but which are broken by exponentially small
non-perturbative effects.  The most natural value for $f_a$ would
appear to be of order $M_{GUT}-M_p$.  Whether this can be made
compatible with cosmology, or whether one can obtain a lower
scale, is an open question.

\section{The Axion in Cosmology}

Despite the fact that it is so
weakly coupled, it be copiously
produced in the early universe\cite{turner}.  The point is the weak coupling
itself.  In the early universe, we know the temperature was once
at lease $1$ MeV, and if the temperature was above a GeV, the
potential of the axion was irrelevant.  Indeed, if the universe is
radiation dominated, the equation of motion for the axion is:
\beq
{d^2 \over dt^2} \phi + {3\over 2 t} {d \over dt} \phi +
V^{\prime}(\phi) =0.
\eeq
For $t^{-1} \gg m_a$, the system is overdamped, and the axion does
not move.  There is no obvious reason that the axion should sit at
its minimum in this early era.  So one can imagine that the axion
sits at its minimum until $t \sim m_a^{-1}$, and then
begins to roll.   For $f_a \sim M_p$, this occurs
at the QCD temperature; for smaller $f_a$ it occurs
earlier.  After this, the axion starts to oscillate in its
potential; it looks like a coherent state of zero momentum
particles.  At large times,
\beq
\phi = {c \over R^{3/2}(t)} \cos(m_a t).
\eeq
so the density is simply diluted by the expansion.  The energy
density in radiation dilutes like $T^4$, so eventually the axion
comes to dominate the energy density.  If $f_a \sim M_p$, the
axion energy density is comparable when oscillations start.
If $f_a$ is smaller, oscillation starts earlier and there
is more damping.  Detailed study (including
the finite temperature behavior of the axion potential) gives a
limit $f_a < 10^{11}$ GeV.

\section{Conclusions:  Outlook}

The strong CP problem, on the one hand, seems very subtle,
but on the other hand it is in many ways similar to the other
problems of flavor which we confront when we examine the standard
model.  $\theta$ is one more parameter which is surprisingly
small.  The smallness of the Yukawa couplings may well be the
result of approximate symmetries.  Similarly, all of the suggestions
we have discussed above to understand the smallness of $\theta$
involve approximate symmetries of one sort or another.

In the case of other ideas about flavor, there is often no
compelling argument for the scale of breaking of the symmetries.
The scale could be so high as to be unobservable, and there is
little hope for testing the hypothesis.
What is perhaps most exciting about the axion is that if we accept
the cosmological bound, the axion might well be observable.

That said, one should recognize that there are reasons to think
that the axion scale might be higher.  In particular, as mentioned
earlier, string theory provides one of the most compelling
settings for axion physics, and one might well expect the
Peccei-Quinn scale to be of order the GUT scale or Planck scale.
There have been a number of suggestions in the literature as to
how the cosmological bound might be evaded in this context.

At a theoretical level, there are other areas in which the
axion is of interest.  Such particles inevitably appear in string
theory and in supersymmetric field theories.  (Indeed, it is
in this context that Peccei-Quinn symmetries of the required
type {\it for} QCD most naturally appear).  These symmetries
and the associated axions are a powerful tool for understanding
these theories.


\noindent
{\bf Acknowledgements:}

\noindent

This work supported in part by a grant from the U.S.
Department of Energy.


\end{document}